# New Families of Large Band Gap 2D Topological Insulators in Ethynyl-Derivative Functionalized Compounds

Lauryn Wu [1,2] *, Kunming Gu [2,3] and Qiliang Li [2] **

[1] Thomas Jefferson High School for Science and Technology, Alexandria, VA 22312, United States of America
[2] Department of Electrical and Computer Engineering, George Mason University, Fairfax, VA 22030, United States of America
[3] Material School of Shenzhen University, Shenzhen Key Laboratory of Advanced Functional Material, Shenzhen, 518060, Guangdong, People's Republic of China

The search for large band gap systems with dissipationless edge states is essential to developing materials that function under a wide range of temperatures. Two-dimensional (2D) topological insulators (TIs) have recently attracted significant attention due to their dissipationless transport, robust properties and excellent compatibility with device integration. However, a major barrier of 2D TIs is their small bulk band gap, which allows for applications only in extremely low temperatures. In this work, first principle calculations were used to analyze the geometric, electronic, and topological properties of $PbC_2X$ and $BiC_2X$ (X = H, Cl, F, Br, I) compounds. The band gap values are remarkably large, ranging from 0.79eV to 0.99eV. The nanoribbons of these compounds exhibited nontrivial topological order in the simulation, thus proving ethynyl-derivative functionalized Pb and Bi films to be new classes of giant band gap 2D TIs. In addition, these findings indicate that chemical functionalization with ethynyl-derivatives is an effective method to tune the band gap and preserve the nontrivial topological order. These novel materials that are applicable at both room temperature and high temperatures open the door to a new generation of electronics.

## I. Introduction

Topological insulators (TIs), with the remarkable quantum spin Hall (QSH) effect, have recently shown great potential in revolutionizing the field of materials science. [1, 2, 3] While behaving like ordinary insulators in the bulk, QSH insulators have metallic states on the edge and are protected by time-reversal symmetry. One of their remarkable characteristics is the absence of backscattering, so electronic currents can flow without dissipation. Furthermore, TIs retain their distinctive properties even when diluted with impurities. Due to their versatility, the QSH insulators are promising in technological applications, such as quantum computing and spintronics. [4]

Despite the fascinating applications of TIs, the lack of large band gap in the bulk is hindering the advancement. Thus far, many three dimensional (3D) compounds, such as $Bi_2Se_3$, $Bi_2Te_3$, and $Sb_2Te_3$, have already been experimentally confirmed to be QSH insulators. [1, 5, 6, 7] On the other hand, the small bulk band gap of TI materials impedes their realistic application. The first material anticipated to be Two dimensional (2D) TI was graphene [8], but weak spin-orbit coupling (SOC) and a small band gap caused TI properties to only be observable in temperatures below 0.01 K. [9, 10] The first experimentally observed and tested 2D TI materials, the quantum wells HgTe/CdTe and InAs/GaSb were similarly observable only at extremely low temperatures. [11, 12, 13] However, compared to 3D materials, 2D TIs have more advantageous features, including better flexibility and being easier to integrate into current electronics. [14] 2D materials can be readily integrated by the well-developed microfabrication technologies for high performance and high-density logic and memory devices. Furthermore, the surface of 3D TIs is not protected against backscattering in any direction other than 180°, whereas 2D TIs have robust edge states that prevent backscattering. [2] In addition, the chemical bonding in 2D structures can be easily modified in post synthesis processes to tune the band gap or achieve certain properties. However, the tiny band gaps in 2D TIs are impeding their progress. Recently, many studies have attempted to investigate 2D TIs with large band gaps for room-temperature applications. [15, 16, 17, 18, 19, 20, 21, 22, 23, 24] Some graphene-like 2D honeycomb structures have been proposed to be QSH insulators, including silicene [25], germanene [26], and stanene. [2] However, the quest for 2D TIs that possess a large band gap is a difficult challenge in order to realize the QSH effect at room temperature.

Several methods can increase the band gap in 2D TI systems, such as placing the materials on a substrate and chemical functionalization. By placing the structure on a substrate, the electronic structure is modified by the interaction between the material and substrate, possibly destroying the topological order. Chemical functionalization of TI monolayers, on the other hand, is a very effective method to widen the band gap and improve structural stability, while preserving the nontrivial topological order. Pristine stanene, for example, has a calculated band gap of 0.1 eV, but with the addition of functional groups, the band gap reached 0.3 eV. [2] Furthermore, the bonds can be easily modified, and therefore the materials can be experimentally realized. [19, 2, 27, 28]

Materials that use heavy atoms, such as Pb and Bi, are favorable in TIs because heavier atoms generally have stronger SOC, leading to a larger band gap. Pb is the heaviest element in group-IV, so it may drive a nontrivial topological order and a large band gap. Despite its huge capability, lead films have not been extensively studied, especially compared to other 2D group-IV honeycomb structures such as graphene and stanene. [29, 2, 30, 31, 8, 32, 33] On the other hand, Bi is the largest element in group-V and has an unusually low toxicity for a heavy metal. It is also known for its strong SOC and its ability to drive a material to a nontrivial topological state. [20] For these reasons, Pb and Bi were chosen to be the base atoms in the study. Moreover, recent studies found that hydrogenation and fluorination of materials have led to quickly increasing lattice disarrangement and defects. [34] Thus, focus has shifted to decorating 2D films with small molecules. [35, 36, 37] Ethynyl-derivatives ($C_2H$, $C_2F$, $C_2Cl$, $C_2Br$, $C_2I$) are excellent options for decorating the surface of 2D structures to enhance the geometric stability, increase the band gap, and preserve a nontrivial topology.

In this work, the structural, electronic, and topological properties of $PbC_2X$ and $BiC_2X$ (X = H, F, Cl, Br, I) monolayers are investigated based on first principle calculations. New classes of QSH insulators in ethynyl-derivative functionalized Pb and Bi monolayers were identified. The analyzed 2D materials achieve substantial band gaps that are large enough to operate in room temperature and high temperature applications. Furthermore, nanoribbon calculations confirm the nontrivial topological order and conducting edge states of the materials. These newly discovered robust TIs are capable for practical use in quantum and electronic devices.

## II. Methods

To investigate the structural and electronic properties, first principle calculations were performed based on the Density Functional Theory (DFT), as implemented in the Virtual Nanolab Atomistix ToolKit (ATK) package. The Perdew-Burke-Ernzerhof (PBE) generalized gradient approximation (GGA) pseudopotential method with the SG15 Optimized Norm-Conserving Vanderbilt (ONCV) basis was used. [38, 39, 40, 41] The mesh energy cutoff was 75 Ha and the total energy convergence criteria was $10^{-6}$ Ha. For unit cell calculations, the integration over the Brillouin zone was sampled with a *$11\times11\times1$* Γ-centered Monkhorst-Pack grid. The nanoribbon calculations were completed with a *$1\times11\times1$* k-mesh. The vacuum region was set to 20 Å to minimize artificial interactions between periodic layers. SOC was considered in the self-consistent calculations. Structural relaxation was performed until the forces on each atom were less than 0.01 eV/Å. The Broyden-Fletcher-Goldfarb-Shanno (BFGS) method was employed for geometry optimization and all structures were fully relaxed.

**III. Results and Discussion**

Figure 1 displays the geometric structure of functionalized Pb and Bi monolayer films. The two center atoms are bonded with ethynyl-derivatives alternating on each side. Like graphene, the hexagonal lattice structure has a threefold rotational symmetry, and therefore has inversion symmetry. However, unlike graphene, the chemically functionalized systems prefer a low buckled configuration, with the two center (Pb or Bi) atoms in the unit cell on different planes.

For each material, a fully relaxed structure was achieved in the calculation, ensuring structural stability for all systems. Table 1 lists the structural parameters, including the optimized lattice constants, buckling heights, bond lengths, and band gaps. The lattice parameter of all the ethynyl-derivative functionalized materials is larger than the pure monolayer. The buckling height for the functionalized Pb films is decreased from the pristine Pb monolayer, but the buckling height increases in the functionalized Bi films, compared to the free-standing Bi monolayer. The bond length between the central atom and $C_2X$ does not have a strong correlation to the X atom, and the bond length in all the Pb systems was 2.22 Å.

To analyze the electronic properties of the systems, band structures with and without SOC were calculated. All Pb based materials in this study have direct band gaps of 0.0 eV without SOC, located at the Γ point. Thus, they all display a gapless semiconductor-like nature in the absence of SOC. This is also referred to as a semimetal, with a degenerate conduction band minimum and valence band maximum at the Fermi level. As confirmed by Figure 2, with SOC, the band gaps open significantly for all the materials in the family, and this can be attributed to the strong SOC of Pb. The SOC effect lifts the energy-degeneracy at the $\Gamma$ point, with the conduction band minimum upshifted and the valence band maximum downshifted. The specific values of the band gap for each system can be found in Table 1. The SOC induced band gap opening at the Fermi level is an indicator for a topologically nontrivial material. Furthermore, in the $PbC_2X$ materials, band inversions are present. Band inversions appear with the inclusion of SOC, which is another strong evidence of nontrivial topology. Without SOC, the band gaps of all the materials are direct, but the band gaps become indirect in the presence of SOC due to the band inversions.

Compared to the free-standing Pb monolayer, the chemically functionalized Pb systems do not have energy-band cross at the Fermi level at the K point, in cases with and without SOC. The functionalized structures also have significantly larger band gaps when SOC is turned on, varying from 0.79eV to 0.87eV, compared to the band gap of 0.37 eV in pure Pb. This shows that chemical functionalization of 2D structures using ethynyl-derivatives is an effective method to tune the band gap. Because these band gap values are massive in comparison to the thermal energy at room temperature ($\approx$ 0.026 eV), ethynyl-derivative functionalized Pb systems are excellent TI material candidates for practical applications.

The band structures of all Bi based monolayers, as shown in Figure 3, were also calculated to examine the electronic and topological properties. Like the Pb film, the Bi monolayer has no band gap, but unlike the Pb monolayer, there is a large gap at the $\Gamma$ point in the Bi band structure. In addition, at the K point, the bands cross twice linearly, once above and once below the Fermi level, in the pure Bi film. With the inclusion of SOC, the energy-degeneracy of the bands around the Fermi level is lifted. Due to the strong SOC of Bi atoms, the band gap of the pure Bi system increases to 0.53eV. The band gap becomes indirect, with the valence band maximum and conduction band minimum near the K point.

Like the functionalized Pb monolayer films, the decorated Bi monolayers have no band gap without SOC. In the absence of SOC, the bands cross linearly once at the K point, forming a Dirac cone with the Fermi level crossing the Dirac point. When SOC is induced, the band gaps at the K point increases considerably. The huge band gaps of the functionalized Pb systems range from 0.93eV to 0.99eV, which exceeds the thermal energy at room temperature. Thus, these materials are predicted to not only be sustainable in room temperature devices, but also viable in high temperature electronics. The specific band gap values can be found in Table 1. Interestingly, with SOC, the CBM is downshifted at the $\Gamma$ point and the VBM remains at the K point, causing an indirect band gap in all $BiC_2X$ materials. Although there is no band inversion in the functionalized Bi systems, the band gap opening in the presence of SOC indicates a TI nature.

The hallmark of 2D QSH insulators is their conducting edge states. The nontrivial topological order of these materials can be proved by the existence of protected gapless edge states. To confirm the nontrivial topology of the systems in the study, the electronic structures of zigzag nanoribbons were calculated. The unit cell of the ribbons was created by transforming the previously calculated slab structures, as seen in Figure 4. The same computational methods used for the bulk structure were employed, with the exception of the k mesh. The vacuum region of 20 Å in the Z direction was maintained, and additional vacuum regions of about 10 Å beside each edge of the nanoribbon were considered to avoid interactions between periodic images. The unit cell of the ribbon had a width of approximately 8nm to prevent interactions between the two edges and all dangling bonds were passivated by hydrogen atoms. The zigzag nanoribbon band structures were calculated with the inclusion of SOC.

The $BiC_2I$ nanoribbon band structure is shown in Figure 4. The the valence and conduction bands are connected, crossing linearly at the $\Gamma$ point. The nanoribbon exhibits a Dirac cone, which verifies the topologically protected conducting edge states. One pair of edge states traverses the bulk band gap, so the valence and conduction bands are connected. Therefore, the Dirac point at the $\Gamma$ point in the $BiC_2I$ nanoribbon signifies that it is a QSH insulator with the key feature of conducting edge states. The presence of the gapless edge states in the nanoribbon agrees with the SOC induced band gap and band inversion in the bulk band structures, proving that these materials are QSH insulators. As shown in Figure 5, the calculated current vs. bias (I-V) curve of the nanoribbon is quite linear with excellent conductivity, further demonstrating the metallic conduction. Such edge states allow for dissipationless electron transport, without backscattering, along the boundary of a material.

**IV. Conclusions**

First principles calculations were performed on $PbC_2X$ and $BiC_2X$ (X = H, Cl, F, Br, I) compounds using DFT. Based on the calculations, we predict new series of 2D TIs and find that chemical functionalization using ethynyl-derivatives is an effective mechanism for tuning the band gap of 2D TIs. The band gap of all functionalized materials, ranging from 0.79eV to 0.99eV, are substantially larger than the band gap of the corresponding pristine monolayer. The largest band gap obtained in this study, 0.99eV, was found in $BiC_2Br$. All studied systems have giant band gap values that are large enough to be sustainable both in room temperature and high temperatures. The SOC induced band gap and band inversion observed in the bulk band structures indicate a nontrivial topological order, and the nanoribbon band structure calculations further confirm the TI properties. The presence of a Dirac point in the bulk gap of the nanoribbon implies conductive, topologically protected edge states and proves a nontrivial phase.

    As an actively researched topic in the recent materials science community, TIs have a huge potential in future technology. Their applications include quantum computers, spintronic devices, thermoelectric devices, infrared detectors, exotic particles, image monopoles, and the realization of Majorana Fermions. The 2D materials in this research have several advantageous characteristics, including easy integration into electronic devices, robust edge states without backscattering, easily modifiable bonds, strong SOC, and huge band gaps. This study on 2D TIs offers new and promising options for dissipationless electronic applications at room temperature.

**Captions:**

Table 1. The lattice constants a (Å), buckling heights h (Å), central atom to $C_2X$ bond lengths d (Å), band gaps without SOC $E_g$ (eV), and SOC induced band gaps $E_{g\text{-SOC}}$ (eV) of all 2D structures in this study.

Figure 1. The hexagonal lattice structure of ethynyl-derivative functionalized 2D systems from the top and side views. The gray atoms represent the center atoms (Pb or Bi); the green atoms represent C, and the red atoms represent H, F, Cl, Br, or I.

Figure 2. Calculated band structures of functionalized Pb films are shown. The first, third, and fifth columns show band structures without SOC, while the second, fourth, and sixth columns show band structures with SOC. The Fermi level is indicated by the dotted green line.

Figure 3. Calculated band structures of functionalized Bi films are shown. The first, third, and fifth columns show band structures without SOC, while the second, fourth, and sixth columns show band structures with SOC. The Fermi level is indicated by the dotted green line.

Figure 4. (a) Calculated zigzag nanoribbon band structure of a 2D $BiC_2I$ monolayer. (b) The zigzag nanoribbon geometric structure that was used to calculate the edge states and current vs. bias characteristics.

Figure 5. The calculated current vs. bias (I-V) curve of the $BiC_2I$ zigzag nanoribbon.

Table

Table 1. The lattice constants a (Å), buckling heights h (Å), central atom to $C_2X$ bond lengths d (Å), band gaps without SOC $E_g$ (eV), and SOC induced band gaps $E_{g\text{-SOC}}$ (eV) of all 2D structures in this study.

| System | a(Å) | h(Å) | d(Å) | $E_g$(eV) | $E_{g\text{-SOC}}$(eV) |
|---|---|---|---|---|---|
| Pb | 5.00 | 1.01 | – | 0.00 | 0.37 |
| $PbC_2H$ | 5.25 | 0.60 | 2.22 | 0.00 | 0.88 |
| $PbC_2F$ | 5.26 | 0.69 | 2.22 | 0.00 | 0.87 |
| $PbC_2Cl$ | 5.27 | 0.58 | 2.22 | 0.00 | 0.79 |
| $PbC_2Br$ | 5.23 | 0.59 | 2.22 | 0.00 | 0.84 |
| $PbC_2I$ | 5.23 | 0.60 | 2.22 | 0.00 | 0.84 |
| Bi | 5.30 | 0.00 | – | 0.00 | 0.53 |
| $BiC_2H$ | 5.54 | 0.01 | 2.22 | 0.00 | 0.96 |
| $BiC_2F$ | 5.54 | 0.06 | 2.20 | 0.00 | 0.93 |
| $BiC_2Cl$ | 5.54 | 0.04 | 2.20 | 0.00 | 0.93 |
| $BiC_2Br$ | 5.52 | 0.10 | 2.20 | 0.00 | 0.99 |
| $BiC_2I$ | 5.52 | 0.08 | 2.12 | 0.00 | 0.97 |

Figures

Figure 1

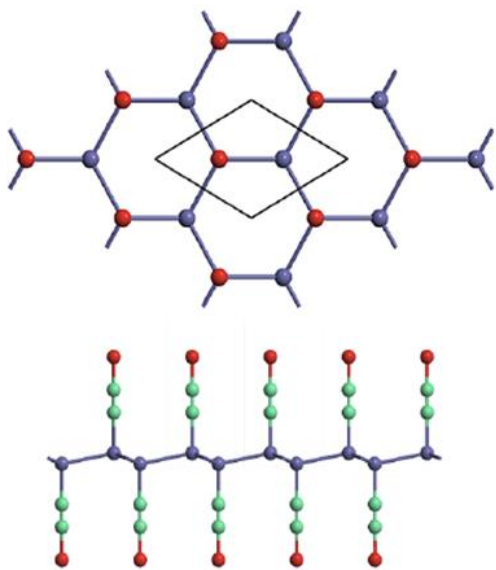

Figure 1. The hexagonal lattice structure of ethynyl-derivative functionalized 2D systems from the top and side views. The gray atoms represent the center atoms (Pb or Bi); the green atoms represent C, and the red atoms represent H, F, Cl, Br, or I.

Figure 2

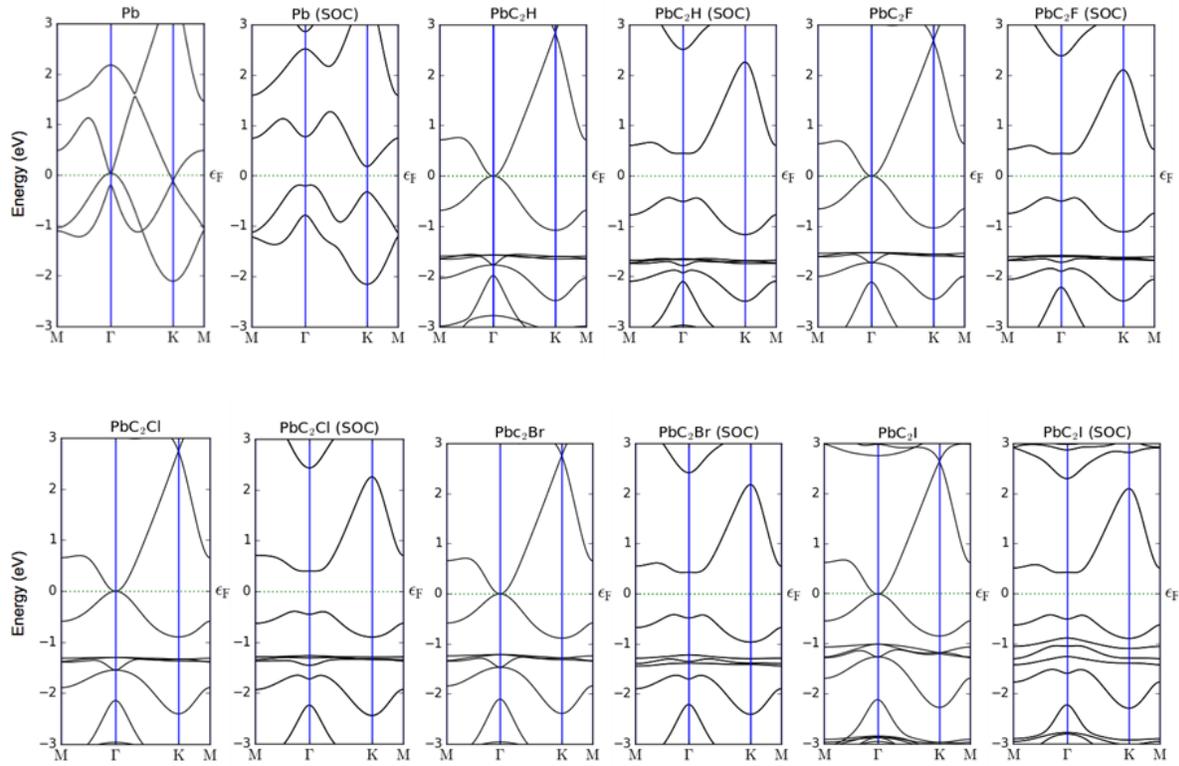

Figure 2. Calculated band structures of functionalized Pb films are shown. The first, third, and fifth columns show band structures without SOC, while the second, fourth, and sixth columns show band structures with SOC. The Fermi level is indicated by the dotted green line.

Figure 3

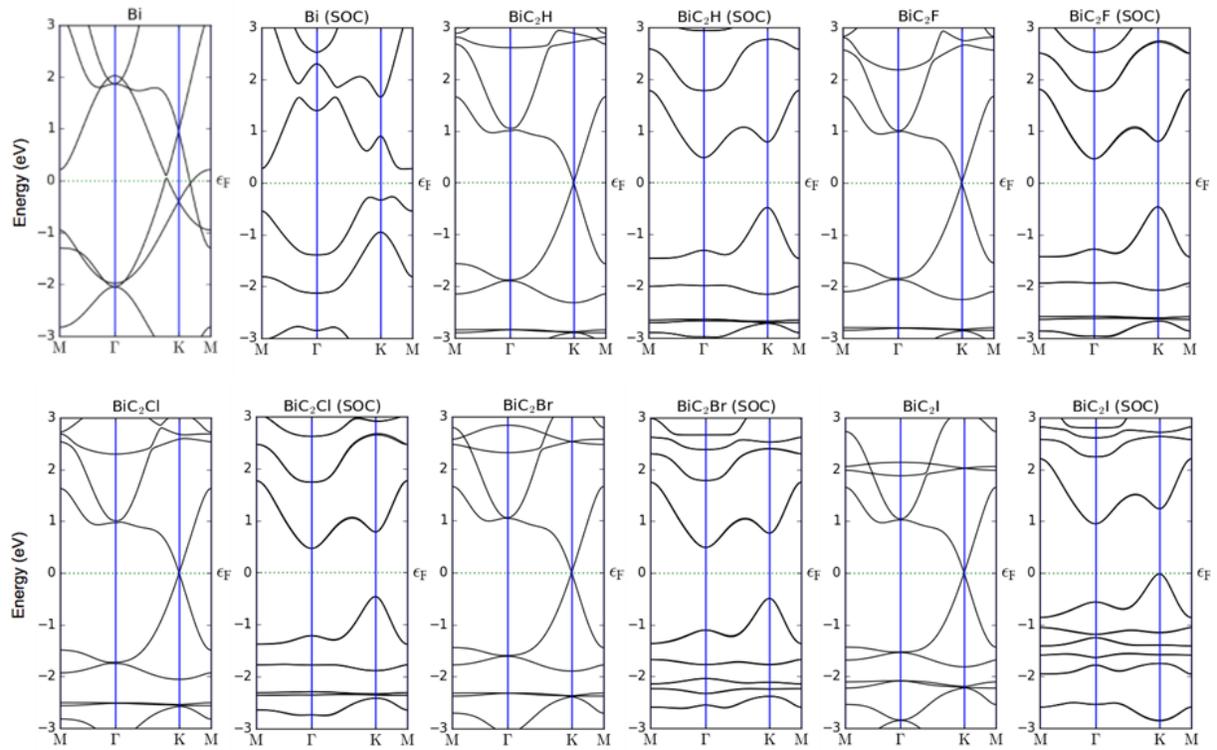

Figure 3. Calculated band structures of functionalized Bi films are shown. The first, third, and fifth columns show band structures without SOC, while the second, fourth, and sixth columns show band structures with SOC. The Fermi level is indicated by the dotted green line.

Figure 4

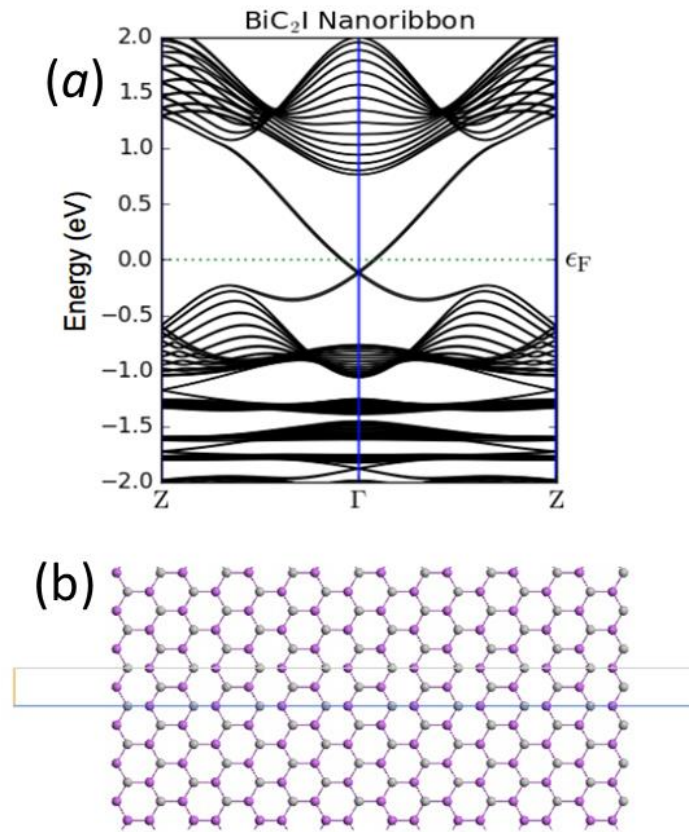

Figure 4. (a) Calculated zigzag nanoribbon band structure of a 2D BiC$_2$I monolayer. (b) The zigzag nanoribbon geometric structure that was used to calculate the edge states and current vs. bias characteristics.

Figure 5

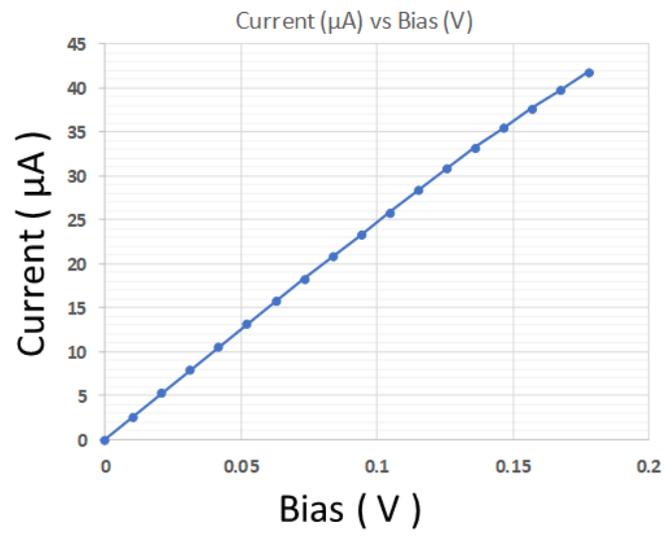

Figure 5. The calculated current vs. bias (I-V) curve of the $BiC_2I$ zigzag nanoribbon.